\documentclass{nature}
\usepackage[utf8]{inputenc}
\usepackage{graphicx}
\makeatletter
\let\saved@includegraphics\includegraphics
\AtBeginDocument{\let\includegraphics\saved@includegraphics}
\renewenvironment*{figure}{\@float{figure}}{\end@float}
\renewcommand*{\fnum@figure}{{\bfseries{}Fig.~\thefigure{}}}  
\renewcommand*{\fnum@table}{{\bfseries{}Table~\thetable{}}}  
\makeatother
\usepackage[]{caption}
\DeclareCaptionLabelSeparator{bar}{\bfseries\ $|$\ }
\usepackage{dcolumn}
\usepackage{bm} 
\usepackage{color,soul}
\definecolor{purple}{RGB}{127,0, 255}

\usepackage{caption}
\usepackage{ragged2e}
\usepackage{amssymb}
\usepackage{amsmath}
\usepackage{mathtools}
\usepackage{multirow}
\usepackage{lineno}
\usepackage{booktabs}

\usepackage{xcolor}
\usepackage[pdftex]{hyperref}
\hypersetup{
  colorlinks,
	linkcolor={blue!75!black!80!yellow},
	citecolor={blue!75!black!80!yellow},
	urlcolor={blue!75!black!80!yellow},
	pdfstartview=FitH
	}

\usepackage{txfonts}
\usepackage{txfontsb}

\usepackage{textcomp} 

\newcommand{\remove}[1]{\textcolor{red!70!black}{\ignorespaces#1}}

\title{Geometric similarities and topological phases in surface magnon polaritons} 

\author{Chen Qian$^{1}$, Jicheng Jin$^{1}$, Thomas Christensen$^{2}$, Li He$^{1}$, Anthony Sigillito$^{3}$, Eugene J. Mele$^{1}$, Bo Zhen$^{1}$}

\begin{document}
\maketitle

\begin{affiliations}
 \item Department of Physics and Astronomy, University of Pennsylvania, Philadelphia, Pennsylvania 19104, USA
 \item Department of Photonics and Electrical Engineering, Technical University of Denmark, Lyngby 2800, Denmark
  \item Department of Electrical and System Engineering, University of Pennsylvania, Philadelphia, Pennsylvania 19104, USA
\end{affiliations}

\section*{Abstract}
Highly spatially-squeezed polaritons, with propagation momentum significantly larger than free-space modes at the same frequency, enable varied and extreme control over light-matter interaction.
Compared to other polaritons, surface magnon polaritons, the magnetic counterpart of surface phonon polaritons, have received relatively little attention. 
Here, we investigate 
the dispersion and properties of surface-magnon polaritons, highlighting the impact of geometric similarities and applying them to various surface-magnon polariton devices in both conventional and topological settings. 
Our theory predicts a method for strongly localizing and significantly enhancing magnetic fields in the microwave range and developing compact and lossless connectors capable of interconnecting waveguides with vastly different input and output impedances.
Our work opens new avenues for manipulating magnetic fields in the microwave regime and for exploring topological phases in polariton platforms.




\maketitle

\section*{Introduction}


The topological phases of electromagnetic waves are an area of recent interest\cite{ozawa2019topological,lu2014topological}.
A characteristic example of such topological phases is the optical quantum anomalous Hall effect\cite{wang2008reflection,haldane2008possible,rechtsman2013photonic}, modeled after Chern insulators. 
Protected by their bulk topology, optical Chern insulators feature unidirectional transport channels at their edges, which are immune to back-scattering due to surface roughness or fabrication imperfections. 
To achieve the Chern insulator phase, breaking time-reversal symmetry (and reciprocity) is a key requirement, which can be implemented by various means, including by the gyromagnetic effect\cite{wang2009observation,skirlo2015experimental, bahari2017nonreciprocal} in ferrimagnetic semiconductors (e.g., yttrium iron garnett, YIG), the gyroelectric effect\cite{haldane2008possible} in magnetized plasmas\cite{prudencio2022ill,qian2023topological}, or in dynamically driven nonlinear materials\cite{he2019floquet,fang2019anomalous}. 
As an important practical matter, the strength of gyrotropic responses is high in the microwave regime but is significantly weakened in the near-infrared or visible regime.
On the other hand, optical nonlinearity exhibits less frequency dependence but requires intense driving fields due to its weak intrinsic strength.
In addition to studying topological phases in pure electromagnetic waves in dispersionless materials, there is a growing interest in exploring topological phases in quasiparticle-settings such as polaritons\cite{basov2016polaritons,he2023polaritonic,karzig2015topological,liu2020generation}, which can exist in systems with strong response dispersion.

Surface polaritons have proven especially effective in controlling light-matter interactions due to their high spatial squeezing factors. 
Compared to free-space plane waves at the same frequency, the spatial oscillation of surface polaritons can be significantly larger. 
Although surface polaritons are subject to material losses\cite{khurgin2015deal}, they support a wealth of interesting phenomena\cite{basov2020polariton}, including strong field enhancements\cite{jornada2020universal}, hyperbolic dispersion\cite{yang2019type,hu2022real,passler2022hyperbolic}, and extreme modifications of the dynamics of nearby emitters\cite{PhysRevResearch.2.023089}, including enabling transitions that are otherwise forbidden\cite{rivera2016shrinking,rivera2020light,sloan2019controlling}.
A particularly interesting regime of surface polaritons is in the quasi-static limit with extremely high spatial squeezing factors. In this regime, the electric ($\mathbf{E}$) and magnetic fields ($\mathbf{B}$) of electromagnetic waves are approximately decoupled, allowing for independent and virtually arbitrary control of both fields. 
Two response types have received the majority of attention in the pursuit of high confinement and squeezing, both involving a negative permittivity $\varepsilon(\omega)<0$, namely response below the plasma frequency in metals\cite{reather1988surface,zayats2005nano,grigorenko2012graphene,garcia2014graphene} (also in indium tin oxide\cite{west2010searching} and gaseous-phase plasma) or within the reststrahlen band of polar crystals\cite{palik1977history}.
Alternatively, materials with negative permeability $\mu(\omega) < 0$, such as those located close to the ferromagnetic resonances in ferrites\cite{fuller1987ferrites},  
can be used, although this approach has received much less attention.  

Here we introduce topological phases into surface-magnon polaritons (SMP) 
by identifying a Chern insulator phase in an array of SMP ring resonators. 
We present a universal method of geometric scaling SMP structures in the quasi-magnetostatic limit, which allows for easy adjustment of their group velocity and effective impedance while keeping their frequencies unchanged. We apply this geometric scaling method to the SMP Chern insulators and their chiral edge states and develop a compact and lossless interconnect between SMP waveguides with vastly different impedances. 
Our work paves the way for future exploration of topological phases in polaritonic systems, controlling (e.g., localizing and enhancing) magnetic fields in the microwave regime, and developing topological microwave devices.

\section*{Main text}
We begin by introducing the geometric scaling rule of electromagnetic waves in the quasi-static limit, before applying it to various structures supporting surface magnon polaritons (SMPs), first in conventional settings (waveguides and ring resonators) and subsequently in topological settings (Chern insulators and chiral edge states).  

\subsection{Geometric scaling of electromagnetic waves in the quasi-static limit}
\begin{figure}[ht]%
\centering
\includegraphics[width=\textwidth]{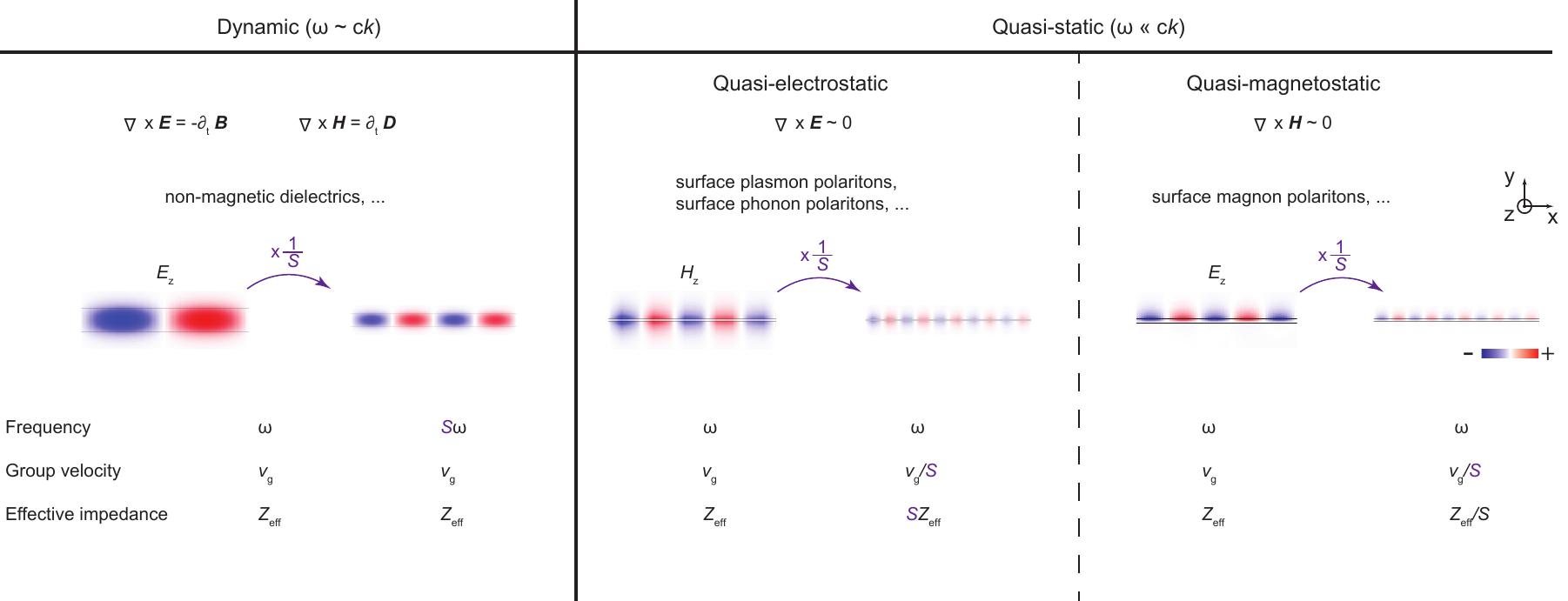}
\caption{\textbf{Geometric scaling of electromagnetic waves in dynamic versus quasi-static settings.} 
Geometrically scaling down a non-magnetic dielectric waveguide by a factor of $s$ in all dimensions increases the operation frequency ($\omega$) by the same factor $s$, while keeping the waveguide group velocity ($v_{\mathrm{g}}$) and effective impedance ($Z_\mathrm{eff}$) unchanged.
On the other hand, geometrically scaling down a surface polariton waveguide, in the quasi-electrostatic or quasi-magnetostatic limit, provides an effective method to adjust the group velocity and the effective impedance, while keeping the operational frequency unchanged. 
}\label{Fig1}
\end{figure}

In a waveguide made of non-magnetic dielectrics (Fig.~\ref{Fig1}), the spatial oscillation frequency of an electromagnetic wave, as characterized by its propagation momentum ($k$), is comparable to its temporal oscillation frequency scaled by the speed of light ($k \sim \omega/c$).
As required by both the (source-free) Faraday law $\bm{\nabla}\times\mathbf{E}=-\partial\mathbf{B}/\partial t$ and the Ampere--Maxwell equation $\bm{\nabla}\times\mathbf{H}=\partial\mathbf{D}/\partial t$ in the vacuum, the electric and magnetic fields are comparable in magnitude ($\textbf{E} \sim Z_0 \textbf{H}$).
Here $Z_0 = \mu_0 c \approx 377\,\Omega$ is the free-space impedance.
In contrast, a different class of solutions to Maxwell's equations exists where the spatial oscillation is significantly faster than the temporal oscillation ($k \gg \omega/c$).
Such solutions, often referred to as surface polaritons, e.g., exist at the interfaces between air and materials with negative permittivity $\varepsilon$ (as in surface-plasmon polaritons\cite{jablan2009plasmonics,grigorenko2012graphene} and surface-phonon polaritons\cite{caldwell2015low}) or negative permeability $\mu$ (surface-magnon polaritons). 
Such surface polaritons are frequently endowed with a very large momentum and an accompanying strong spatial in- and out-of-plane localization (or squeezing), reflecting their substantial coupling to a matter component.
As a result, in strongly localized surface-plasmon polaritons (surface-magnon polaritons), the magnetic (electric) field strength nearly vanishes: i.e., 
$\mathbf{B} \propto \mathbf{H} \sim \bm{0}$ ($\mathbf{D} \propto \mathbf{E} \sim \bm{0}$).
To still satisfy the Faraday  law (Ampere--Maxwell equation) despite the vast difference in field magnitudes, the corresponding $\mathbf{E}$ 
($\mathbf{H}$)
field must be nearly curl-free and can be expressed as the gradient of a scalar potential, similar to the electrostatic (magnetostatic) fields at zero frequency.

Electromagnetic fields scale very differently in dynamic settings compared to quasi-static settings.
When a structure consisting of a dispersionless medium is shrunk by a factor of $s$ in all dimensions (Fig.~\ref{Fig1}), the eigenfrequencies of any modes supported by structure increase by a factor of $s$ from $\omega$ to $s\omega$; as does the spatial oscillation frequency (e.g., momenta increase from $k$ to $sk$).
Meanwhile, the electric and magnetic fields, $\textbf{E}$ and $\textbf{H}$, scale in the same way, leaving the effective impedance $Z_{\mathrm{eff}}$ unchanged, since $|\mathbf{E}|\sim Z_{\mathrm{eff}}|\mathbf{H}|$.
The group velocity $v_{\mathrm{g}} = \partial\omega/\partial k$ is similarly invariant in the dispersionless case.  
In contrast, the $\mathbf{E}$ and $\mathbf{H}$ fields are effectively decoupled in the quasi-static limit, allowing them to vary independently under scaling transformations.
For example, when a polaritonic waveguide is shrunk by a factor of $s$, the operating frequency $\omega$ remains unchanged.
Meanwhile, only the $\mathbf{E}$ ($\mathbf{H}$) field is increased by a factor of $s$ in the quasi-electrostatic (quasi-magnetostatic) setting, while the complementary fields $\mathbf{H}$ ($\mathbf{E}$) remains nearly unchanged.
The effective impedance $Z_{\mathrm{eff}}$ is thus increased (decreased) by a factor of $s$ while the group velocity $v_{\mathrm{g}}$ is reduced by a factor of $s$.


\subsection{SMP waveguides}
\begin{figure}[ht]%
\centering
\includegraphics[width=0.5\textwidth]{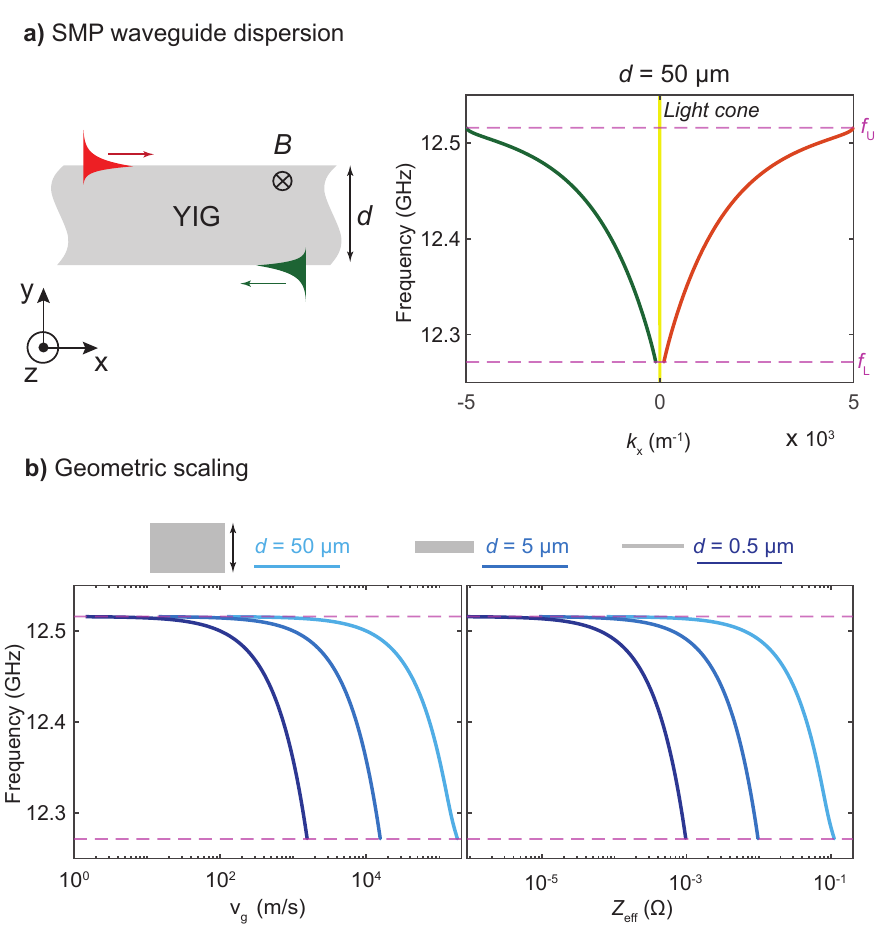}
\caption{\textbf{SMP waveguide dispersion and geometric scaling.} 
\textbf{a}, Schematic drawing of a surface magnon polariton (SMP) waveguide based on ferrimagnetic semiconductors (YIG) placed in an external magnetic field. The two SMP modes (red and green) live on opposite surfaces, and travel in opposite directions, but are both confined between frequencies $f_{\rm U}$ and $f_{\rm L}$.
\textbf{b}, Scaling down the waveguide width $d$ by factors of $s=10$, from 50 to 5 and then to $0.5\,\text{\textmu{}m}$, results in a proportional decrease in the group velocity $v_{\mathrm{g}}$ and effective impedance $Z_{\rm eff}$.
}\label{Fig2}
\end{figure}

Motivated by these simple considerations, we derive the surface-magnon polariton (SMP) waveguide dispersions and show how they change under geometric scaling.
An SMP waveguide consists of a finite-width ferrimagnetic semiconductor (e.g., YIG) placed in an external magnetic field, 
$B_{\rm ext}\hat{\mathbf{z}}$, directed tangentially to the waveguide's surfaces which are assumed to span the $xz$ plane (Fig.~\ref{Fig2}a).
The YIG magnetic permeability tensor\cite{pozar2011microwave} can be expressed as
{
\renewcommand\arraystretch{.65}
\begin{align}
\boldsymbol{\mu} &= \boldsymbol{\eta}
\begin{pmatrix}
\mu_{+} &  & \\
 & \mu_{-} & \\
 &   & \mu_0 
\end{pmatrix}
\boldsymbol{\eta} ^{\dagger}, 
\\
\boldsymbol{\eta} &=  
\begin{pmatrix}
\frac{1}{\sqrt{2}} & \frac{\mathrm{i}}{\sqrt{2}} & \\
\frac{\mathrm{i}}{\sqrt{2}} & \frac{1}{\sqrt{2}} & \\
 &   & 1
\end{pmatrix}\!. 
\end{align}%
}%
Here the column vectors of the unitary matrix $\boldsymbol{\eta}$ label the optical principle axes $\hat{\mathbf{e}}_{\pm} = (\hat{\mathbf{x}} \pm \mathrm{i}\hat{\mathbf{y}})/\!\sqrt{2}$ and $\hat{\mathbf{e}}_3 = \hat{\mathbf{z}}$ of the magnetized YIG.
Along the principal axes $\hat{\mathbf{e}}_{\pm}$, the permeability is:  
\begin{equation} \label{magetized-dispersion}
    \mu_{\pm}
    =
    \mu_0 (1+\chi_\pm)
    =
    \mu_0 \left[1 + \frac{f_{\rm M}}{(f_{\rm H} \pm f)+\mathrm{i}\alpha f} \right]. 
\end{equation}
Here $f=\omega/2\pi$ is the frequency, $f_{\mathrm{H}} = \gamma_0 H$, $f_{\mathrm{M}} = \gamma_0 M$, and $\gamma_0 = 2.8\,\text{MHz/oe}$ is the gyro-magnetic ratio of YIG.  In all calculations below, we take the external magnetic field
$H=B_{\rm ext}/\mu_0=3570\,\text{oe}$\remove{,} and the magnetization of YIG as its saturated magnetization $M=1800\,\text{oe}$, following Refs.~\citenum{pozar2011microwave} and \citenum{carsten2017sub}.
Meanwhile, the Gilbert damping constant is taken to be $\alpha = 8.9\times10^{-4}$, consistent with recent experiments\cite{kosen2019microwave,pirro2021advances}.
In finite 3D YIG samples, the build-up of a demagnetization field also affects the permeability\cite{pozar2011microwave, gurevich1996magnetization}: in the present 2D context, however, the demagnetization field is zero since the sample extends infinitely in the direction of the external magnetic field. 
See Methods for more details on the material properties of YIG.   
In addition to a continuum of bulk electromagnetic modes, a set of localized SMP modes are found at the interfaces between YIG and air (Fig.~\ref{Fig2}a). 
These modes are also known as magnetostatic surface waves (MSSW)\cite{stancil2012theory,gurevich1996magnetization}, which are usually derived under the magnetostatic approximation ($\textbf{E}=\bm{0}$). 
Here, we solve the full Maxwell's equations without such approximations throughout, which leads to a few important deviations from the conventional MSSW results shown later.

The SMP waveguide modes are bounded in frequency, lying between a lower limit of $f_{\rm L} = \sqrt{f_{\rm H} (f_{\rm H} + f_{\rm M})}$ and an upper limit of $f_{\rm U} = f_{\rm H} + f_{\rm M}/2$.
The frequency $f_{\rm L}$ is often denoted by $f_\perp$ in the MSSW literature\cite{gurevich1996magnetization}, as it represents the resonance frequency of a forward volume wave with magnetic field polarized perpendicularly to the external magnetic field, i.e., with $\mathbf{B}\perp \mathbf{B}_{\mathrm{ext}}$.
For frequencies outside this range, there exists a continuum of bulk modes below $f_{\rm L}$ but no modes with $\mathbf{B}\perp \mathbf{B}_{\mathrm{ext}}$ above $f_{\rm U}$.
We note that our calculated SMP dispersion starts from a finite momentum at the lower frequency bound $f_{\rm L}$, as opposed to starting from zero momentum as in the MSSW literature derived under the magnetostatic approximation\cite{stancil2012theory,gurevich1996magnetization}. 
The SMP dispersion consists of two branches (Fig.~\ref{Fig2}a): in one branch (red curve) the mode is localized on the upper interface and travels to the right; in the other (green curve), it is localized on the bottom interface, traveling leftward.
Such waveguide modes are non-reciprocal in nature, consistent with the fact that both time-reversal symmetry and reciprocity are broken by the YIG permeability.
The SMP dispersion exhibits a $\omega(k_x) = \omega(-k_x)$ symmetry because the two SMP branches are related by the $C_2^z$ symmetry of the waveguide. 
The momentum of both SMP branches, $k_x>10^3\,\text{m}^{-1}$, is significantly larger than that of free-space modes at the same frequencies ($f/c\sim40\,\text{m}^{-1}$), confirming the quasi-magnetostatic nature of the SMP modes.

Next, we geometrically scale SMP waveguides to different widths ($d$) and confirm the scaling rules of SMP modes.
As the waveguide's width ($d$) is shrunk by factors of $s=10$, from $50\,\text{\textmu{}m}$ to $5\,\text{\textmu{}m}$ to $0.5\,\text{\textmu{}m}$, the frequency range of SMP remains unchanged, while the in-plane momentum $k_x$ increases by factors of 10 (Fig.~\ref{Fig2}b).
Accordingly, the group velocity $v_{\mathrm{g}}$ and the effective impedance $Z_{\rm eff}$ also decrease by factors of $s=10$, consistently with the predictions above.

\subsection{SMP ring resonators}
\begin{figure}[ht]%
\centering
\includegraphics[width=0.5\textwidth]{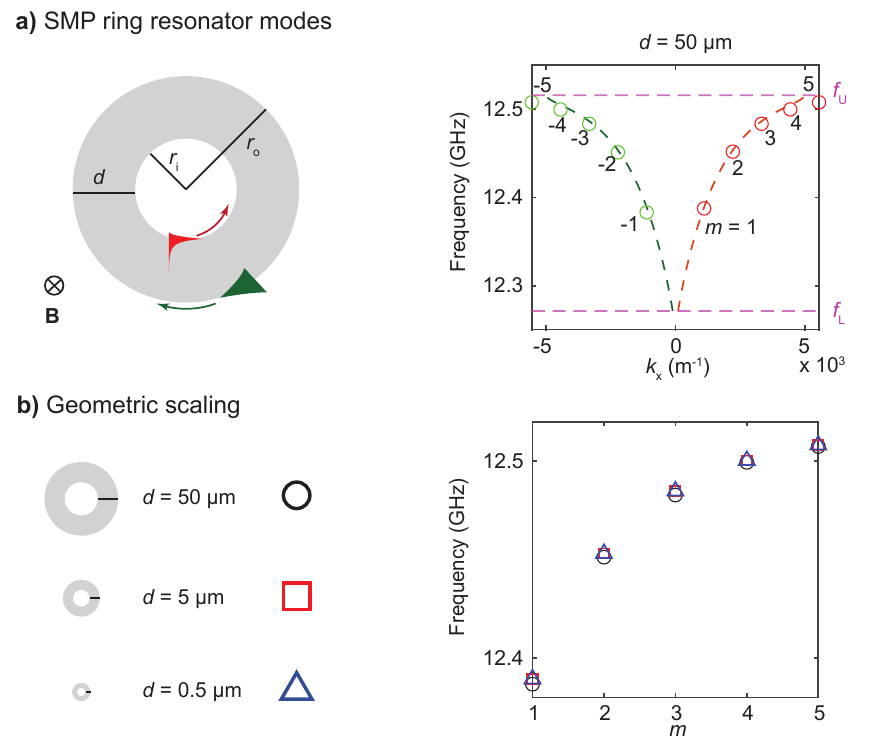}
\caption{\textbf{SMP ring resonators and scaling invariance. } 
\textbf{a}, SMP ring resonators support two sets of resonances, labeled by positive and negative azimuthal numbers $m$, that propagate on opposite surfaces and in opposite directions. The resonance frequencies of SMP agree well with the waveguide dispersion of SMP with the same width $d$. 
\textbf{b}, Geometrically scaling down the SMP ring resonators by factors of $s=10$ does not alter the resonance frequencies.
}\label{Fig3}
\end{figure}

SMP resonances can be built by bending an SMP waveguide into a ring resonator with inner radius $r_{\rm i}$, outer radius $r_{\rm o}$, and width $d = r_{\rm o} - r_{\rm i}$ (Fig.~\ref{Fig3}a).
The resulting rotational symmetry quantizes the SMP ring resonances according to their azimuthal numbers $m$: $\textbf{E}_{z}^{m} \propto \mathrm{e}^{\mathrm{i} m \phi}$, where $\phi$ is the polar angle in cylindrical coordinates. 
Similarly to the SMP waveguide modes, the cylindrical SMP resonances with $m>0$ ($m<0$) live at the inner (outer) surface and travel in the counter-clockwise (clockwise) direction. 
The SMP resonance frequencies (circles) are in good agreement with the SMP waveguide dispersion (dashed lines) when associating the momentum of the SMP resonance with the azimuthal number $m$ according to $k= |m|/r_{\rm eff}$, where $r_{\rm eff}=d/\!\ln(r_{\rm o}/r_{\rm i})$, as demonstrated in the example with $d=50\,\text{\textmu{}m}$ and $r_{\rm i}=120\,\text{\textmu{}m}$.
See Supplementary Information for more details.
As a result, the SMP resonance frequency increases with $\lvert m \rvert$. 
Our calculations indicate that SMP resonances with opposite $m$ are very close in frequency, but they are not identical as previously noted in the MSSW literature\cite{stancil2012theory,gurevich1996magnetization}.
Similar geometric scaling behavior can be observed in SMP ring resonances (Fig.~\ref{Fig3}b): as the entire structure is scaled down by a factor of 10 (100), the SMP resonance frequencies labeled by red squares (blue triangles) remain nearly identical to their original values labeled as black circles.
 
\subsection{SMP Chern insulators and chiral edge states} 
\begin{figure}[ht]%
\centering
\includegraphics[width=0.5\textwidth]{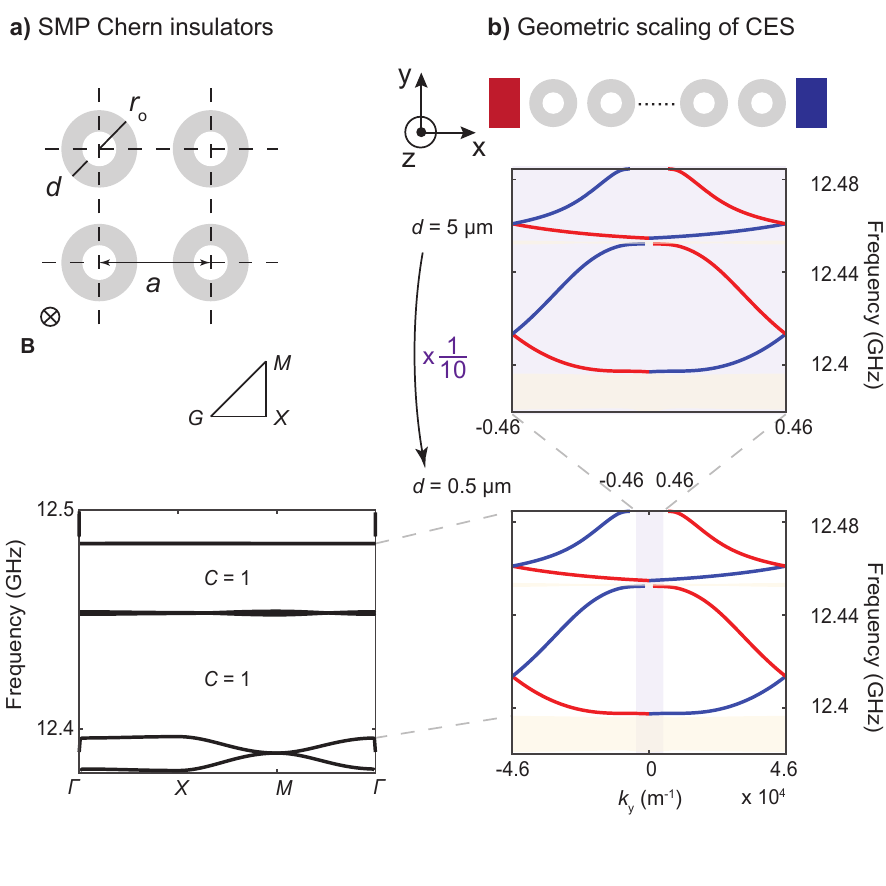}
\caption{\textbf{SMP Chern insulators and geometric scaling of chiral edge states.} 
\textbf{a},~Schematic of a square lattice array of SMP ring resonators (width $d=5\,\text{\textmu{}m}$) with a periodicity of $a=108\,\text{\textmu{}m}$. The band structure features a number of topological energy gaps with non-zero Chern numbers $C=1$.
\textbf{b},~The bulk topology is confirmed by the existence of chiral edge states at interfaces between a ring-resonator super-cell and perfect electric conductors (red and blue rectangles). Scaling down the Chern insulator unit cell by a factor of $s=10$ (reducing $d$ from $5\,\text{\textmu{}m}$ to $0.5\,\text{\textmu{}m}$) keeps the frequency of chiral edge states unchanged while increasing the momentum $k_y$ by a factor of $s=10$.
}\label{Fig4}
\end{figure}

A Chern insulator made of SMPs can be created by arranging SMP ring resonators in a square lattice array (as shown in Fig.~\ref{Fig4}a).
Each ring in the lattice is identical to the one shown in Fig.~\ref{Fig3}b, with a width of $d=5\,\text{\textmu{}m}$. When the lattice constant is $a=108\,\text{\textmu{}m}$, the band structure of this SMP ring resonator array consists of nearly dispersionless bands, which are built from ring resonator modes with opposite $m$ (such as $m=\pm2$ or $m=\pm3$). The only exception is the three highly dispersive bands, which are mostly built from $m=0$ and $m=\pm1$ modes.
Many of the SMP energy gaps are topologically non-trivial, characterized by non-zero Chern numbers (e.g., $C=1$), which is consistent with the analysis of $C_2^z$ indices of the eigenmodes at high symmetry points ($\Gamma$ and $M$) in the band structure. See Supplementary Information for more details. To confirm the Chern numbers of the SMP energy gaps, a super-cell geometry is constructed to check for unidirectional chiral edge states (CESs) at the interfaces between SMP ring resonators and perfect electric conductors (PECs) (Fig.~\ref{Fig4}b). 
Indeed, pairs of CESs are found between 12.397 and 12.452\,GHz and also between 12.455 and 12.483\,GHz. These pairs of CESs, labeled by red and blue lines, localize at opposite interfaces (left and right) and travel in opposite directions (downward and upward). We note that compared to previous photonic Chern insulators that also use YIG\cite{wang2008reflection,wang2009observation,skirlo2015experimental}, our unit cells and feature sizes are about 2--3 orders of magnitude smaller, which confirms the high spatial squeezing factors of our SMP. 

Both the SMP Chern insulator and CES can also be geometrically scaled: when the unit cell is scaled down by a factor of $s=10$, reducing $d=5\,\text{\textmu{}m}$ to $0.5\,\text{\textmu{}m}$ and reducing $a=108\,\text{\textmu{}m}$ to $10.8\,\text{\textmu{}m}$, the band Chern numbers remain the same and the CES frequencies remain unchanged (Fig.~\ref{Fig4}b, bottom panel), but now with 10 times the momentum $k_y$. Accordingly, the waveguide group velocity $v_{\mathrm{g}}$ and effective impedance $Z_{\rm eff}$ are also reduced by factors of $s=10$.

\subsection{Topological interconnects between SMP waveguides with arbitrary effective impedance} 
\begin{figure}[ht]%
\centering
\includegraphics[width=\textwidth]{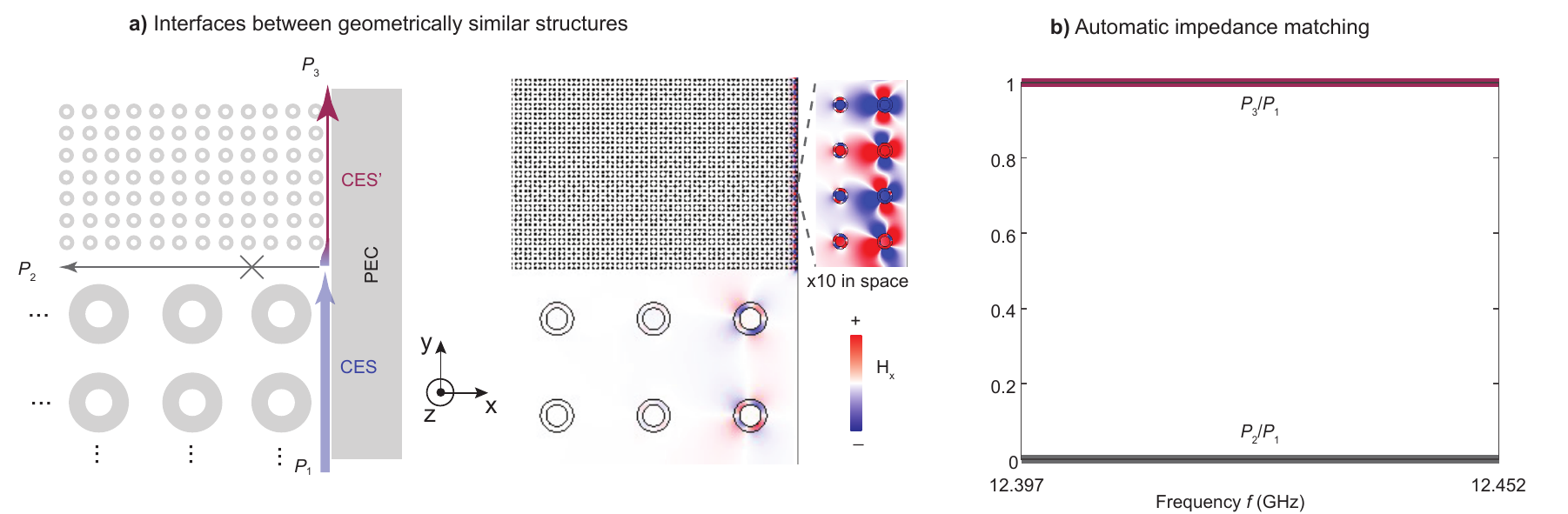}
\caption{%
    \textbf{Topological interconnect between SMP waveguides of different impedance.} 
    \textbf{a},~Schematic of an interconnect between an SMP chiral edge state (CES, blue arrow at the bottom) and its geometrically scaled-down version (CES$'$, purple arrow at the top).
    \textbf{b},~Input power from the bottom ($P_1$) is almost perfectly converted into the top CES mode ($P_3/P_1\sim 1$) provided that the trivial edge state (gray arrow) is eliminated by design ($P_2 \sim 0$). 
}\label{Fig5}
\end{figure}

A compact and lossless interconnect can be constructed between geometrically similar SMP chiral edge states. 
Specifically, an SMP Chern insulator (the bottom section of Fig.~\ref{Fig5}a) is interfaced with a geometrically scaled-down version of itself (by a factor of $s=20$, the top section), both terminated by a PEC on the right. 
Following the rules of geometrical scaling (Fig.~\ref{Fig1}), the two SMP Chern insulators, with bigger and smaller unit cells, share the same frequencies, as do the CES at their interfaces with the PEC on the right. 
For an input power of $P_1$ into the bottom CES (blue  arrow), due to its unidirectional nature, there are only two possible output channels. 
It can travel through the scaled-down version chiral edge state (purple arrow, labeled as CES$'$) to the top ($P_3$), which has a significantly reduced group velocity and effective impedance. 
Otherwise, it has to propagate through trivial edge states  (gray arrow) at the interface between two Chern insulators of the same Chern numbers towards the left ($P_2$). 
By engineering the interface configuration, one can eliminate such trivial edge states and thus ensure that $P_2 \sim 0$ and $P_3/P_1 \sim 1$ (Fig.~\ref{Fig5}b). 
See Supplementary Information for more details on trivial edge state engineering. 
Overall, we have developed a lossless and compact interconnect between SMP waveguides with significantly different effective impedance and group velocities, based on topology and geometric similarity. 

\section*{Discussion and conclusion}

Our proposed structures can be readily fabricated and tested in experiments. To begin with, though our calculations are for 2D YIG structures, similar results can also be achieved in 3D structures defined in thin-film YIG and placed in between two metallic plates---a similar geometry as in previous experimental demonstrations\cite{wang2009observation,skirlo2015experimental}.
Furthermore, thin-film YIG wafers are readily available with narrow ferromagnetic resonance linewidths\cite{carsten2017sub} (down to 1.5oe). Standard etching mechanisms have also been reported in the literature\cite{9706176,zhu2022inverse} to define proposed structures such as YIG waveguides and ring resonators. 
Our derived geometric scaling rules pose a few limitations on the geometric parameters in practice, as summarized in Table.~E1. On the one hand, very wide waveguides (e.g., $d\gg 100\,\text{\textmu{}m}$) do not produce enough spatial squeezing to be in the quasi-magnetostatic limit and the SMP dispersion becomes insensitive to the change of $d$. On the other hand, in extremely narrow waveguides (e.g., $d\ll 1\,\text{\textmu{}m}$), the magnetic dipole exchange interaction can no longer be neglected, as we have assumed in our calculations.
Further investigation into such small structures allows exploration into the non-local regime of electromagnetic waves\cite{PhysRevB.96.104441, yang2019general}, where more complex and intriguing phenomena may be discovered.  

In summary, we introduce topological phases into surface magnon polaritons through the example of Chern insulators. By applying the geometric scaling rules in the quasi-magnetostatic regime, we demonstrate a powerful mechanism to adjust the speed of light and the effective impedance of SMP waveguides. Based on this scaling mechanism, we also develop a topological and lossless interconnect between SMP waveguides with vastly different impedance. 
Our work opens up new possibilities in exploring topological phases in polaritonic systems, manipulating magnetic fields in the microwave regime, and developing compact topological devices for microwave applications.

\section*{Methods}
\subsection{Permeability and permittivity of YIG} The ferromagnetic resonance (FMR) of YIG is at frequency $f=f_{\rm H}$. 
Close to the FMR, $\mathop{\mathrm{Re}}\mu_-$ drops sharply to a negative value, and it reaches $-1$ at the frequency of $f_{\rm U}=f_{\mathrm{H}}+f_{\mathrm{M}}/2$, which sets the upper frequency bound for SMP. 
Meanwhile, if transformed back into the Cartesian coordinate, the diagonal terms of the permeability matrix read $\mu_x=\mu_y=(\mu_++\mu_-)/2$, which follows the general trend of $\mu_-$ near the FMR. Meanwhile $\mu_{x,y}$ reach 0 at $f_{\rm L}=\sqrt{f_{\mathrm{H}}(f_{\mathrm{H}}+f_{\mathrm{M}})}$, which defines the lower frequency bound for SMP. 
Taken together, SMPs are confined to the frequency regime of above $f_{\rm L}$ ($\mu_y > 0$) and below $f_{\rm U}$ ($\mu_{-}<-1$), as shown in Fig.~E1. 
Out of the two loss mechanisms in YIG, magnetic and electric, in highly squeezed SMP with much stronger magnetic fields than electric fields, the magnetic loss (Gilbert damping) dominates over the electric (dielectric loss tangent).
The Gilbert damping constant $\alpha$ of YIG varies strongly with temperature and is taken to be $8.9\times 10^{-4}$ in our calculations, which is at the higher end of reported values\cite{jermain2017increased,kosen2019microwave}.  
The permittivity of YIG is taken to be $\epsilon=15$, while the dielectric loss tangent (nominally around $2\times 10^{-4}$ in the literature\cite{how2000single}) is neglected in our calculations as it is dominated by the Gilbert damping constant in the quasi-magnetostatic limit.




\subsection{Data availability} The data within this paper are available from the corresponding author upon request. 

\subsection{Acknowledgments} The authors acknowledge stimulating conversations with Nicholas Rivera, Jamison Sloan, and Marin Solja\v{c}i\'c. 
This work was partly supported by the U.S. Office of Naval Research (ONR) through grant N00014-20-1-2325 on Robust Photonic
Materials with High-Order Topological Protection and grant N00014-21-1-2703, the Air Force Office of Scientific Research through grant FA9550-21-1-0299. Work by T.C. is supported by the Villum Fonden (42106). Work by E.J.M is supported by the Department of Energy under grant DE-FG02-84ER45118.

\subsection{Author Contributions} C.Q. and B.Z. conceived the project. C.Q. performed numerical simulations assisted by J.J. C.Q. and B.Z.
wrote the paper with input from all authors. All authors discussed the results. B.Z. supervised the project. 

\subsection{Competing interests} The authors declare no competing interest.

\subsection{Correspondence} Correspondence should be addressed to B.Z. (email: bozhen@sas.upenn.edu).

\clearpage

\makeatletter
\renewcommand*{\fnum@figure}{{\bfseries{}Extended Data Fig.~E\thefigure{}}}  
\renewcommand*{\fnum@table}{{\bfseries{}Extended Data Table~E\thetable{}}}  
\makeatother
\setcounter{figure}{0} 
\setcounter{table}{0} 

\begin{figure}[ht]%
\centering
\includegraphics[width=0.5\textwidth]{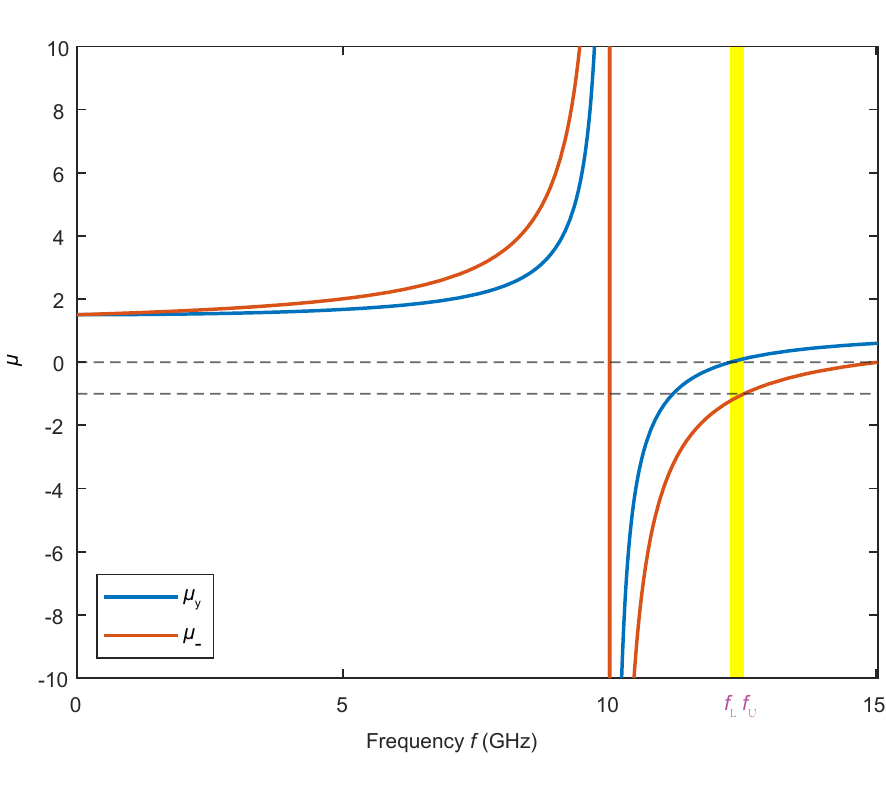}
\caption{%
    \textbf{Permeability dispersion of YIG and frequency bounds of SMP.} 
    Surface magnon polaritons are bound in the frequency regime to be above $f_{\mathrm{L}}$ ($\mu_y > 0$) and below $f_{\mathrm{U}}$  ($\mu_{-}<-1$). 
    \label{FigE1}
    }
\end{figure}


\clearpage

\begin{table} 
\centering
\small\addtolength{\tabcolsep}{1pt}
\setlength{\arrayrulewidth}{1pt}
\footnotesize	
\begin{tabular}{rccc}
\toprule
 & Quasi-static \& Dipolar regime \\ 
\midrule
Waveguide width $d$ (\textmu{}m)  & $50$\,--\,$0.5$\\
Momentum $k$ (m$^{-1}$) &  $\sim 10^3$\,--\,$10^5$ \\
Group velocity $v_{\mathrm{g}}$ (m/s) &  $\sim 10^5$\,--\,$10^3$ \\
Impedance $Z_{\rm eff}$ ($\Omega$) & $\sim 10^{-1}$\,--\,$10^{-3}$ \\
\bottomrule
\end{tabular}
\caption{
    \textbf{Eligible parameter regime for SMP waveguides.}
    Our discussion of SMP and the geometric scaling rule is bound in the parameter space (e.g., waveguide width $d$): on the one hand, very wide waveguides (e.g., $d\gg 100\,\text{\textmu{}m}$) break the quasi-magnetostatic limit. On the other hand, magnetic dipole exchange interaction cannot be neglected in very narrow waveguides (e.g., $d\ll 1\,\text{\textmu{}m}$), causing ferromagnetic resonance frequency dispersion that is neglected in our current calculations.
    \label{TabE1}
    }
\end{table}

\clearpage

\section*{References}
\bibliographystyle{naturemag}
\bibliography{maintext.bib}

\end{document}